\documentclass[reprint,nofootinbib,superscriptaddress,amsmath,amssymb,aps]{revtex4-1}
\usepackage{graphicx}
\usepackage{rotating}
\usepackage{dcolumn}
\usepackage{bm}
\usepackage{color}
\usepackage{mathptmx, textcomp}
\usepackage[latin1]{inputenc}
\usepackage{braket}
\usepackage[colorlinks,linkcolor=magenta,anchorcolor=cyan,citecolor=blue]{hyperref}
\usepackage[multiple]{footmisc}

\def\be{\begin{equation}}
\def\ee{\end{equation}}
\def\ba{\begin{eqnarray}}
\def\ea{\end{eqnarray}}

\def\nn{\nonumber}

\newcommand{\eq}[1]{(\ref{#1})}

\def\n{\nonumber}\def\q{\theta}      \def\p {\pi} \def\a {\alpha} \def\s {\sigma} \def\d {\delta} \def\f {\phi} \def\g {\gamma} \def\h {\eta}   \def\l {\lambda} \def\z {\zeta} \def\x {\xi} \def\c {\chi} \def\b {\beta}  \def\m {\mu} \def\pd {\partial}\def\p {\pi}   \def \e { \varepsilon}
      \def\S {\Sigma}   \def\G {\Gamma}     \def\grad{\nabla}\def\.{\cdot}
\def\math {\mathcal}
\begin{document}

\title{Weak Cosmic Censorship Conjecture of Hairy Black Holes in Einstein Gravity}
\author{Aofei Sang}
\email{aofeisang@mail.bnu.edu.cn}
\affiliation{College of Education for the Future, Beijing Normal University, Zhuhai 519087, China}
\affiliation{Department of Physics, Beijing Normal University, Beijing 100875, China\label{addr2}}
\author{Jie Jiang}
\email{Corresponding author. jiejiang@mail.bnu.edu.cn}
\affiliation{College of Education for the Future, Beijing Normal University, Zhuhai 519087, China}
\affiliation{Department of Physics, Beijing Normal University, Beijing 100875, China\label{addr2}}
\date{\today}

\begin{abstract}

Apart from the Kerr-Newman black holes, the hairy black holes in general relativity have been widely investigated in the gravity and cosmology.
In this paper, we extend the Sorce-Wald method to prove the weak cosmic censorship conjecture (WCCC) of the static and spherically symmetric hairy black holes in the Einstein gravity without using the explicit expressions of the metric and lagrangian for matter fields. We examine the WCCC in the collision process without requiring the spherically symmetry of the perturbation matter fields. After assuming the stability condition of spacetime and applying the Gaussian null coordinates into the variational identities, we derive the first two order perturbation inequalities which reflects the null energy condition of the matter fields. As a result, we find that nearly extremal static hairy black hole cannot be destroyed in the above perturbation process under the second-order approximation. Our result implies that the valid of the WCCC is universal for the hairy black hole in the Einstein gravity as long as the matter fields satisfy the null energy condition under the perturbation level.

\end{abstract}
\maketitle
\section{Introduction}
The weak cosmic censorship conjecture (WCCC) \cite{Penrose:1969pc, Wald:1997wa} is one of the unproved conjectures in classical general relativity. It is proposed by Penrose \cite{Penrose:1969pc} and states that there is no naked singularity in our universe. This conjecture ensures that the spacetime singularity caused by the gravitational collapsing body is always hidden inside the event horizon of the black hole. It also means that any physical process cannot destroy the black hole and cause the naked singularity. To examine the WCCC, Wald proposed a gedanken experiment in 1974 \cite{Wald:1974wl}, in which they considered a test particle throwing into an extremal Kerr-Newman (KN) black hole and showed that any particles which can enter the horizon cannot destroy the black hole under first-order approximation. After that, Hubeny \cite{Hubeny:1998ga} started with a nearly extremal black hole and found that there exists some situation under the second-order approximation where the black hole can be destroyed if the test particle satisfies some certain conditions. Since then, Hubeny-type violations were found in many black holes and gravitational theories \cite{Jacobson:2010iu,Chirco:2010rq,Saa:2011wq,Gao:2012ca}. However, this method viewed the black hole as a background and does not consider the second-order effects, such as the self-force effect, radiative and finite-size effects \cite{Jacobson:2010iu}. Later, numerical simulation considering the self-force effect on particles indicates the Hubeny-type violation will not happen \cite{Zimmerman:2012zu,Colleoni:2015ena,Barausse:2011vx,Barausse:2010ka}.

In 2017, Sorce and Wald \cite{Sorce:2017dst} put forward a new version of gedanken experiment. By assuming the stability condition and the null energy condition of the matter field, they derived the first-order and second-order perturbation inequalities using the variational identities. Based on these inequalities, they proved that a nearly extremal KN black hole cannot be over-charged or over-spun under the second-order approximation of perturbation. In this method, they consider the full dynamical process of both the spacetime and the matter fields, and therefore self-force effect, finite-size effect, and other second-order corrections are automatically taken into account. Recently, Sorce-Wald gedanken experiments are widely investigated in various black hole models and gravitational theories.

The no-hair theorem states that a stationary black hole in Einstein-Maxwell theory is only determined by the mass, electric charge, and angular momentum \cite{Israel:1967wq, Carter:1971zc, Ruffini:1971bza}, i.e., it is described by the KN black hole solutions. However, the hairy black holes can be found when the gravity couples to some extra field degree of freedom, such as the Yang-Mills fields \cite{MSDV, Bizon:1990sr, Greene:1992fw}, dilaton filed \cite{Kanti:1995vq}, and Skyrme hairs \cite{Luckock:1986tr,Droz:1991cx}. Most recently, by considering the complex scalar field without the static symmetry, Herdeiro and Radu \cite{Herdeiro1,Herdeiro2,Herdeiro:2015waa} presented a family of spinning scalar-hairy black hole solutions in Einstein gravity minimally coupled to a massive complex scalar field.  Moreover,  Hong, Suzuki, and Yamada \cite{Hong:2020miv} also showed that there exists static and spherical charged scalar hair after the scalar mass term is considered. Additionally, the Astro observation \cite{Planck:2015fie} implies that there exists dark matter \cite{Bertone:2004pz} and dark energy \cite{Peebles:2002gy} in real universe. It is natural to ask whether the WCCC is valid in these hairy black holes. In this paper, we would like to extend the Sorce-Wald method to prove the WCCC of static hairy black holes for the Einstein gravity coupled to some additional matter fields without using the explicit expressions of the metric and Lagrangian for matter fields.

The outline of this paper is as follows. In Sec. \ref{sec2}, we consider the collision process in the static and spherically symmetric hairy black in Einstein gravity couples to some additional matter fields. Then, we assume that the spacetime satisfies the stability condition and introduce two coordinate systems to describe this assumption. In Sec. \ref{sec3}, we review the Noether charge method of Iyer and Wald and derive the first two order off-shell variational identities in Einstein gravity. In Secs. \ref{sec4} and \ref{sec5}, after applying the Gaussian null coordinate of the hypersurface into the variational identities, we derived the first- and second-order perturbation inequalities from the null energy condition. Then, we prove that the nearly extremal static hairy black hole cannot be destroyed under the above collision process. Finally, the conclusion and discussion are presented in Sec. \ref{sec6}.

\section{Perturbed static black hole geometries}\label{sec2}

In this paper, we consider a static hairy black hole in the Einstein gravity coupled to some matter fields, such as the electromagnetic field, scalar field, dilaton field, dark matter, and dark energy. The Lagrangian $n$-form of this theory is given by
\ba\begin{aligned}
\bm{L}=\frac{\bm{\epsilon}}{16\p} R+\bm{L}_\text{mt}\,,
\end{aligned}\ea
in which $R$ is the Ricci scalar, $\bm{\epsilon}$ is the volume element of the metric $g_{ab}$, and $\bm{L}_\text{mt}$ is the Lagrangian of the matter fields. The equation of motion of this theory is given by
\ba
G_{ab}=8\p T_{ab}\,,
\ea
where $G_{ab}=R_{ab}-1/2Rg_{ab}$ is the Einstein tensor, and $T_{ab}$ is the total energy stress tensor of the matter fields.

Then, we take a general ansatz for the static and spherically symmetric solution with the line element
\ba\label{metback}
ds^2=-f(r)dv^2+2\chi(r) dv dr+r^2 h_{ij}d\q^i d\q^j\,,
\ea
with  $i,j=1,2\dots,n-2$ in the ingoing Eddington's coordinate $O: \{v, r, \q\}$, in which $h_{ij}d\theta^i d\theta^j$ is the line element of the unit sphere, and $\chi(r)$ is a positive function of $r$. If there exists a positive $r_h$ such that $f(r_h)=0$, the above solution describes a black hole. In this case, the black hole possess an event horizon at $r=r_h$. If there is no positive root of $f(r)=0$, the geometry describes a naked singularity.

Next, to test whether the black hole can be destroyed, we consider a process where some matter falls into the black hole from a finite portion of the horizon and finally, the spacetime settles down to a static state, i.e. we assume that the black hole satisfies the stability condition \cite{Hollands:2012sf,Sorce:2017dst}. This means, at a very early time, the spacetime geometry is the same as the background spacetime, i.e. there exists a coordinate system $O_1: \{v_1, r_1, \q_1\}$ such that the metric has the same form with Eq. \eqref{metback}, i.e.,
\ba\label{metricO1}
ds^2=-f(r_1)dv^2_1+2\c(r_1) dv_1 dr_1+r^2_1 h_{ij}d\q^i_1 d\q^j_1\,.
\ea

The stability condition implies that there exists another coordinate system $O_2: \{v_2, r_2, \q_2\}$ such that the late-time geometry can be described by the line element
\ba\label{metricO2}
ds^2=-f_2(r_2)dv^2_2+2\c_2(r_2) dv_2 dr_2+r^2_2 h_{ij}d\q^i_2 d\q^j_2\,.
\ea
The form of $f_2(r)$ and $\c_2(r)$ are determined by the collision process. Note that $O_1$ and $O_2$ are two separated coordinate systems and both of them cannot cover the whole spacetime. With these setups, examining the WCCC converts to checking whether $f_2(r)$ possesses a positive root.

Now, we consider a one-parameter family $\f(\l)$ of the field configurations \cite{Sorce:2017dst}, in which every element $\f(\l)$ represents a process described above and $\f(0)$ is the background static geometry with the line element \eq{metback}. When $\l$ is a small parameter, the configuration $\f(\l)$ describes a perturbation process on $\f(0)$. In the background geometry, both the coordinate systems $O_1$ and $O_2$ become $O$. The early-time geometry is given by Eq. \eqref{metricO1} in the coordinate $O_1$ and the corresponding metric component is independent on $\l$. Noting that The dependence on $\lambda$ of the components show the different influences of the collision process, the late-time geometric function $f_2(r)$ and $\c_2(r)$ are determined by $\l$. Therefore, the line element of the late-time geometry in $O_2$ can be shown by
\ba\label{metriclatetime}
ds^2(\lambda)=-f(r_2, \l)dv^2_2+2\c(r_2, \l) dv_2 dr_2+r^2_2 h_{ij}d\q^i_2 d\q^j_2\,.\quad
\ea
Under the background spacetime with $\l=0$, we have $f(r, 0)=f(r)$ and $\c(r, 0)=\c(r)$. The above setups indicate examining the WCCC is equivalent to finding the root of $f(r, \l)$.

\section{Off-shell variational identity in Einstein gravity}\label{sec3}

In this section, we would like to review the Noether current method proposed by Iyer and Wald \cite{Iyer:1994ys} and derive the off-shell variational identities. In Einstein gravity, the gravitational part of the Lagrangian $n$-form is
\ba\begin{aligned}
\bm{L}_\text{grav}=\frac{\bm{\epsilon}}{16\p}R\,.
\end{aligned}\ea
Variation of the Lagrangian gives
\begin{equation}\label{fdeltalag}
	\delta \boldsymbol{L}_\text{grav} = \boldsymbol{E}_g^{ab} \delta g_{ab} + d \boldsymbol{\Theta}(g, \delta g)\,,
\end{equation}
where
\ba\begin{aligned}\label{egdeltag}
\boldsymbol{E}_\text{grav}^{ab} = -\frac{\boldsymbol{\epsilon}}{16\p} G^{ab}=-\frac{\bm{\epsilon}}{2}T^{ab}\,.
\end{aligned}\ea
describes the gravitational part of the equation of motion, and
\ba\begin{aligned}\label{theta}
\boldsymbol{\Theta}_{a_2\dots a_n} (g, \delta g) = \frac{1}{16 \pi} \bm{\epsilon}_{da_2\dots a_n} g^{de} g^{fg} \left(\nabla_g \delta g_{ef} - \nabla_e \delta g_{fg} \right)\,.\quad
\end{aligned}\ea
is the symplectic potential $(n-1)$-form. Here we define
\ba\begin{aligned}
\d^k \h(x)=\left.\frac{\partial^k\h(x,\lambda)}{\partial\lambda^k}\right|_{\l=0}\,.
\end{aligned}\ea
to denote the $k$th-order variation of the quantity $\h(x, \l)$. Note that the symbol $\delta$ always denotes the partial derivative respect to $\lambda$ when we fix the variable in the other slot of $\h$. We can define the symplectic current $(n-1)$-form as
\begin{equation}\label{definitionomega}
	\boldsymbol{\omega} (g, \delta_1 g, \delta_2 g) = \delta_1 \boldsymbol{\Theta} (g, \delta_{2} g) - \delta_{2} \boldsymbol{\Theta} (g, \delta_{1} g)\,,
\end{equation}
which can be explicitly written as
\begin{equation}
	\omega_{a_2\dots a_n} = \frac{1}{16 \pi} \bm{\epsilon}_{da_2\dots a_n} w^d,
\end{equation}
where
\begin{equation}
	w^a = P^{abcdef} \left(\delta_{2} g_{bc} \nabla_d \delta_{1} g_{ef} - \delta_1 g_{bc} \nabla_d \delta_{2}g_{ef} \right),
\end{equation}
with
\ba\begin{aligned}	
P^{abcdef} &= g^{ae} g^{fb} g^{cd} - \frac{1}{2} g^{ad} g^{be} g^{fc} - \frac{1}{2} g^{ab} g^{cd} g^{ef}\\
& - \frac{1}{2} g^{bc} g^{ae} g^{fd} + \frac{1}{2} g^{bc} g^{ad} g^{ef}.
\end{aligned}\ea
The Noether current $(n-1)$-form $\boldsymbol{J}_\zeta$ associated with the vector field $\zeta^a$ is given by
\begin{equation}\label{fjexp}
	\boldsymbol{J}_\zeta = \boldsymbol{\Theta} (g, \math{L}_\z g) - \zeta \cdot \boldsymbol{L}.
\end{equation}
On the other hand, the Noether current can also be written as \cite{Iyer:1994ys}
\begin{equation}\label{sjexp}
	\boldsymbol{J}_\zeta = \boldsymbol{C}_\zeta + d \boldsymbol{Q}_\zeta
\end{equation}
with the Noether charge
\ba\begin{aligned}\label{chargegr}
	\left(\bm{Q}_\zeta \right)_{a_3\dots a_n} = - \frac{1}{16 \pi} \bm{\epsilon}_{a_3\dots a_ncd} \nabla^c \zeta^d
\end{aligned}\ea
and the constraints
\ba\begin{aligned}
	(\bm{C}_{\zeta})_{a_2\dots a_n}=2\bm\epsilon_{ea_2\dots a_n}\zeta^a G_a^e\,.
\end{aligned}\ea
Then, combining the above expressions and replacing $\z^a$ by the Killing vector filed $\x^a$ in the background geometry $\f(0)$, when $\xi^a$ is fixed under the variation, we can further obtain the first-order variational identity\cite{Sorce:2017dst}
\ba\begin{aligned}\label{variation1}
&\quad d\left[\delta \boldsymbol{Q}_\x - \z \cdot \boldsymbol{\Theta} \left(g, \delta g \right)  \right]+\x \cdot \boldsymbol{E}_g^{ab} \delta g_{ab} + \delta\boldsymbol{C}_\x=0
\end{aligned}\ea
and second-order variational identity
\ba\begin{aligned}\label{variation2}
&\quad d\left[\delta^2 \boldsymbol{Q}_\x - \z \cdot \d\boldsymbol{\Theta} \left(g, \delta g \right)  \right]\\
& = \boldsymbol{\omega} \left(g, \delta g, \mathcal{L}_\x \d g \right)- \x \cdot \boldsymbol{E}_g^{ab} \delta g_{ab} - \delta\boldsymbol{C}_\x\,.
\end{aligned}\ea
It is worth noting that the validity of these variational identities is independent of the gravitational theory imposed in spacetime.

\section{First-order Sorce-Wald Gedanken experiment}\label{sec4}
In this section, we consider the first-order Sorce-Wald gedanken experiment in the above perturbation process.
To test the WCCC, we need to know whether there exists an event horizon after perturbation, which is equivalent to checking whether the root of $f(r,\lambda)=0$ exists. Then, it is natural to convert this question to check whether the minimum value of $f(r,\lambda)$ is negative. Thus, we define a function
\be\begin{aligned}
h(\lambda)=f(r_m(\lambda),\lambda),
\end{aligned}\ee
where $r_m(\lambda)$ denotes the radius where $f(r_m(\lambda),\lambda)$ is the minimum value of $f(r,\lambda)$ and $r_m(\lambda)$ can be determined by $\partial_r f(r_m(\lambda),\lambda)=0$. For a nearly extremal black hole, we can define a small parameter $\epsilon$ such that $r_m=(1-\epsilon)r_h$ with $r_h$ satisfied $f(r_h)=f(r_h,0)=0$. Then $h(\lambda)$ can be expanded as
\be\begin{aligned}\label{flambda}
h(\lambda)&=f(r_m)+\lambda\delta f(r_m)\\
&+\frac{\lambda^2}{2}\left[\delta^2 f(r_m)+2\delta r_m\delta f'(r_m)+\delta r_m^2 f''(r_m) \right]+O(\lambda^3)\,,
\end{aligned}\ee
Considering $f'(r_m)=0$, we have
\be\begin{aligned}
  f'(r_h)=\epsilon r_h f''(r_h)+\math{O}(\epsilon^2)
\end{aligned}\ee
For the first term of Eq. \eqref{flambda}, replacing $r_m$ by $(1-\epsilon)r_h$, expanding to the second order of $\epsilon$ and , we can find
\be\begin{aligned}
  f(r_m)=-\frac{1}{2}\epsilon^2 r_h^2 f''(r_h)+O(\epsilon^3)
\end{aligned}\ee
Considering these conditions, we can find that the final expression of $h(\lambda)$ should be
\ba\begin{aligned}\label{hlexp}
h(\lambda)=&\lambda\delta f(r_h)-\frac{1}{2}\epsilon^2 r_h^2 f''(r_h)-\lambda\epsilon r_h \delta f'(r_h)\\
&+\frac{\lambda^2}{2}\left[\delta^2 f(r_h)+2\delta r_m\delta f'(r_h)+\delta r_m^2 f''(r_h) \right]\\
&+\math{O}(\lambda^3,\epsilon^3,\dots)
\end{aligned}\ea

If the order of $\lambda$ is higher than $\epsilon^2$, the leading term of $h(\l)$ is $-1/2\epsilon^2 r_h^2 f''(r_h)$, which is negative. Thus the black hole cannot be destroyed in this case. If $\lambda$ has the same order with $\epsilon^2$, we can find that the black hole is much easier to be destroyed when $-1/2\epsilon^2 r_h^2 f''(r_h)$ is small. Thus, we consider the case where $\lambda$ and $\epsilon$ have the same order.

Note that the coordinate $O_1$ and $O_2$ are two separated coordinate systems and therefore they cannot be used to describe the collision process. To obtain a coordinate system to describe the process, we first construct a Gaussian null coordinate near a hypersurface $\S$ which is the horizon $r=r_h$ in the background geometry. Moreover, it is given by $r_1=r_h$ in $O_1$ at the early time as well as $r_2=r_h$ in $O_2$ at the late time. For now, the choice of its middle part outside $O_1$ and $O_2$ is arbitrary. Here $r_h$ is only the horizon radius of the background geometry and it is independent of the variational parameter $\l$, i.e., it is not the root of the blackening factor $f(r, \l)$.

Next, we would like to introduce the Gaussian null (GN) coordinate $O_\text{GN}$, $\{u,z,\theta^i\}$, near the hypersurface $\Sigma$. In Gaussian null coordinate system,
\ba
k^a=\left(\frac{\partial}{\partial u}\right)^a
\ea
is the tangent vector of $\Sigma$. In the background, the hypersurface is the horizon, which is a null hypersurface and thus $k^ak_a=0$ on the background. In this coordinate system, coordinates $\theta^i$ on $\Sigma$ are carried out by $k^a$ from one cross-section to another on $\Sigma$. And the coordinates outside the surface $\Sigma$ is carried out by $l^a=(\pd/\pd z)^a$, which is a null vector field satisfying $l^a k_a=1$ on $\S$ and $z$ is an affine parameter. i.e., we have
\ba\begin{aligned}
l^a\grad_a l^b=0\,.
\end{aligned}\ea
Moreover, we also assume that $l^a$ is normal to the cross-section of $\S$, i.e., we have
\ba
l^a\left(\frac{\pd}{\pd \q^i}\right)_a=0\,.
\ea
With the above preparations, in the GN coordinate, the line element in the neighborhood of $\Sigma$ can always be written as \cite{Hollands:2012sf}
\ba\begin{aligned}\label{gaussian}
  ds^2=2(dz-\alpha du-\b_i d\q^i)du+\g_{ij} d\q^id\q^j,
\end{aligned}\ea

Next, we would like to find the line element in the Gaussian null coordinate system at the late time. We assume there is a point $p$ on $\Sigma$ with coordinates $(v_0,r_h,\theta^i_0)$ in coordinate system $O_2$, in which $x^\m_2$ denotes the coordinates of $O_2$. Denote $x^{\mu}_2=\tilde{x}^\m(z)$ is the integral curve of $l^a$. This integral curve connects $p$ with another point $p'$ with coordinates $(v_0,z,\theta^i_0)$ in $O_{\text{GN}}$ near $\Sigma$. Meanwhile, for a point near $\Sigma$, the coordinates of $p'$ in $O_2$ can be expressed as
\ba\begin{aligned}
x{}^\m_2=x^{\m}_0+z\left.\frac{d \tilde{x}^\m}{d z}\right|_p+\frac{z^2}{2}\left.\frac{d^2 \tilde{x}^\m}{d z^2}\right|_p+\math{O}(z^3)\,.
\end{aligned}\nn\\\ea
Since $l^a$ is a null vector and satisfies $l^ak_a|_{\Sigma}=0$, we can find
\ba\begin{aligned}
l^a|_{\Sigma}=\frac{1}{\chi(r_h,\lambda)}\left(\frac{\partial}{\partial r}\right)^a
\end{aligned}\ea
Using $l^a\nabla_al^b=0$, we can further obtain
\ba
\left.\frac{d^2 \tilde{x}^\m}{dz^2}\right|_{p}=-\left.\G^\m{}_{\nu\s}l^\nu l^\s\right|_{p}\,.
\ea
Then, the coordinate transformation between $O_2$ and $O_{\text{GN}}$ can be written as
\ba\begin{aligned}
r_2&=r_h+\frac{z}{\chi(r_h,\lambda)}-\frac{z^2\chi'(r_h,\lambda)}{2\chi(r_h,\lambda)^3}\\
v_2&=v_0\,,\quad \quad\theta_2^i=\theta_0^i\,,
\end{aligned}\ea
where we denote
\ba\begin{aligned}
\chi'(r_h,\lambda)=\left.\frac{d\chi(r,\lambda)}{dr}\right|_{r_h}
\end{aligned}\ea
We can further obtain the line element in $O_{\text{GN}}$
\ba\begin{aligned}\label{gaussianlatetime}
ds^2=2(dz-\alpha du)du+\g_{ij} d\q^id\q^j\,,
\end{aligned}\ea
where
\ba\begin{aligned}
\alpha&=\frac{1}{2}f(r_h,\lambda)+\frac{z f'(r_h,\lambda)}{2\chi(r_h,\lambda)}\\
\g_{ij}&=r_h\left(r_h+\frac{2z}{\chi(r_h,\lambda)}\right)h_{ij}\,,
\end{aligned}\ea

Next, we choose a gauge such that the GN coordinate system is fixed under the variation. Taking $\zeta^a=\left({\partial}/{\partial u}\right)^a$ in the first-order variational identity \eqref{variation1} and integrating it over $\Sigma$, we have
\ba\begin{aligned}\label{first}
  \int_{B_1}\left[\delta \bm Q_{\zeta}-\zeta\cdot\bm \Theta(g,\delta g)\right]=-\int_{\Sigma}\delta\bm C_{\zeta},
\end{aligned}\ea
After considering $f(r_h)=0$ on background and using the line element \eq{gaussianlatetime} of late-time geometry in GN coordinate, we can find the explicit expression of the left side of Eq. \eqref{first},
\ba\begin{aligned}
  \int_{B_1}\left[\delta \bm Q_{\zeta}-\zeta\cdot\bm \Theta(g,\delta g)\right]=\frac{-\delta f(r_h)}{8\pi r_h \chi(r_h)} A_H\,,
\end{aligned}\ea
where $A_H$ is the area of the horizon in the background geometry. The right side of Eq. \eqref{first} is
\ba\begin{aligned}
  \int_{\Sigma}-\delta\bm C_{\zeta}=2\int_{\Sigma}\hat{\bm\epsilon}\left(dz\right)_e\left(\frac{\partial}{\partial u}\right)^a\delta T_a^e\,,
\end{aligned}\ea
where $\hat{\bm\epsilon}$ is the induced volume element on $\Sigma$. We assume the perturbation matter field satisfies the null energy condition ,i.e. $T_{ab}(\lambda)k^a(\lambda)k^b(\lambda)\geq 0$ for any null vector $k^a$. We choose the null vector
\ba\begin{aligned}
  k^a(\lambda)=\left(\frac{\partial}{\partial u}\right)^a+\alpha(\lambda)\left(\frac{\partial}{\partial z}\right)^a
\end{aligned}\ea
Using the line element \eqref{gaussian}, we also have the matrix of the inverse metric is
\ba\begin{aligned}
&g^{zu}=1\,,\qquad g^{zi}=-\gamma^{ij}\b_j\,,\\
&g^{zz}=2\alpha+\gamma^{ij}\b_i\b_j\,,\qquad g^{ij}=\gamma^{ij}\,,
\end{aligned}\ea
in which $\gamma^{ij}$ is the inverse of $\gamma_{ij}$.
Then, it is not hard to verify
\ba\begin{aligned}\label{energycondition}
  T_{ab}k^ak^b=T_u^z-\gamma^{ij}\b_i\b_j T_{uz}+\alpha^2 T_{zz}+\gamma^{ij}\b_j T_{ui}\,.
\end{aligned}\ea
For simplification, we have neglected $\l$ in the above expression. From the above relation, we can get
\ba
\d (T_{ab}k^ak^b)=\d T_u^z\,,
\ea
Here we have used the fact that $\a=\b_i=T_{ui}=0$ under the background geometry. And therefore the null energy condition under the first-order approximation implies
\ba
\d T_u^z\geq 0\,.
\ea
Then, the variational identity reduces to
\ba
\d f(r_h)\leq 0\,.
\ea
Then, Eq. \eq{hlexp} reduces to
\ba\begin{aligned}
h(\l)\leq \math{O}(\l^2, \e^2, \l \e)\,,
\end{aligned}\ea
which means $f(r,\lambda)$ exists a positive root under the first-order approximation and therefore the black hole cannot be destroyed under the first-order approximation. However, there exists an optimal condition $\delta f(r_h)=0$ such that the leading term of $h(\l)$ is second order, which means that we cannot determine the sign of $h(\l)$ under the first-order approximation in this case and we need to consider the second-order approximation. From the above derivation, it is not hard to see that the optimal condition $\delta f(r_h)=0$ also implies
\ba\begin{aligned}
\delta T_{kk}=\delta T_u^z=0\,.
\end{aligned}\ea

\section{Second-order Sorce-Wald gedanken experiment}\label{sec5}
In this section, we consider the second-order Sorce-Wald gedanken experiment when the optimal condition is satisfied. Note that the choice of the middle of the hypersurface $\S$ is arbitrary in the last section. In the appendixes, we have shown that we can choose the hypersurface such that it is a null hypersurface when $\e=0$ under the first-order approximation of $\lambda$, i.e., we have
\ba\label{alphabeta}
\a =z \tilde{\a}+\math{O}(\e \l, \l^2, \e^2)\,,\quad \b_i =z \tilde{\b}_i+\math{O}(\e \l, \l^2, \e^2)\,,
\ea
and the optimal condition can also give
\ba
\d T_u^z=\d T_{uu}=0
\ea
on this new hypersurface, in which the variation is evaluated in its GN coordinate system.

Next, we consider the second-order variation identity
\be\begin{aligned}\label{firstidentity}
  &d\delta\left[\delta \bm Q_{\zeta}-\zeta\cdot\bm \Theta(g,\delta g)\right]\\
  &=\bm \omega(g;\delta g,\mathcal{L}_{\zeta}\delta g)-\zeta\cdot\delta(\bm E_{g}\delta g)-\delta^2 \bm C_{\zeta}\,,
\end{aligned}\ee
and integrate it over the null hypersurface described above. Noting $\zeta^a=\left({\partial}/{\partial u}\right)^a$ is tangent to $\Sigma$, we can get
\be\begin{aligned}
  &\int_{B_1}\delta\left[\delta \bm Q_{\zeta}-\zeta\cdot\bm \Theta(g,\delta g)\right]\\
  &=\int_{\Sigma}\bm \omega(g;\delta g,\mathcal{L}_{\zeta}\delta g)-\int_{\Sigma}\delta^2 \bm C_{\zeta}\,,
\end{aligned}\ee
We can use the metric \eqref{gaussian} to calculate the left side directly. After considering the optimal condition $\delta f (r_h)=0$, we can find
\be\begin{aligned}
  \int_{B_1}\delta\left[\delta \bm Q_{\zeta}-\zeta\cdot\bm \Theta(g,\delta g)\right]=-\frac{\delta^2 f(r_h)}{8\pi r_h h(r_h)}A_H\,.
\end{aligned}\ee
Expanding \eqref{energycondition} to the second order and using the first order optimal condition $\d (T_{ab}k^ak^b)=\d T_u^z=0$, \eqref{energycondition} gives
\be\begin{aligned}
\delta^2(T_{ab}k^ak^b)=\delta^2 T_u^z+\mathcal{O}(\e) \geq 0
\end{aligned}\ee
Thus,
\be\begin{aligned}
  -\int_{\Sigma}\delta^2 \bm C_{\zeta}=\int_{\Sigma}\hat{\epsilon}\delta^2 T_u^z\geq \mathcal{O}(\lambda)
\end{aligned}\ee
Next, we define the canonical energy of the perturbation $\delta g$ on $\Sigma$ by
\be\begin{aligned}
\math{E}_\S(g,\d g)\equiv \int_{\Sigma}\bm\omega(g;\delta g,\mathcal{L}_{\xi}\delta g)\,.
\end{aligned}\ee
The canonical energy can be obtained as
\ba\begin{aligned}
&\math{E}_\S(g,\d g)\\
&=\int_{\Sigma}\bm{\epsilon}_{aa_1\dots a_{n-1}}P^{abcdef}\left(\mathcal{L}_{\xi}\delta g_{bc}\nabla_d\delta g_{ef}-\delta g_{bc}\nabla_d \mathcal{L}_{\xi}\delta g_{ef}\right)\\
&=-2\int_{\Sigma}\tilde{\bm\epsilon}P^{zbcdef}\mathcal{L}_{\xi}\delta g_{bc}\nabla_d\delta g_{ef}+\int_{\Sigma}\mathcal{L}_{\xi}(\tilde{\bm\epsilon}P^{zbcdef}\delta g_{bc}\nabla_d \delta g_{ef})
\end{aligned}\n\\\ea
Using the Stokes' theorem, considering the assumption that the perturbation vanishes on $B_0$ and $\delta g_{ab}=\delta \g_{ab}+\mathcal{O}(\epsilon)$ on $\Sigma$ according to \eqref{alphabeta}, we can get
\ba\begin{aligned}
\math{E}_\S&(g,\d g)\\
=&-2\int_{\Sigma}\tilde{\bm\epsilon}P^{zbcdef}\mathcal{L}_{\xi}\delta \g_{bc}\nabla_d\delta g_{ef}+\int_{B_1}\hat{\bm\epsilon}P^{zbcdef}\delta \g_{bc}\nabla_d \delta g_{ef}\\
&+\math{O}(\e)
\end{aligned}\n\\\ea
Using the late-time metric \eqref{metriclatetime}, it is not hard to verify the boundary term vanishes. Thus, we have
\ba\begin{aligned}
  \math{E}_\S(g,\d g)=-2\int_{\Sigma}\tilde{\bm\epsilon}P^{zbcdef}\mathcal{L}_{\xi}\delta \g_{bc}\nabla_d\delta g_{ef}+\math{O}(\e)
\end{aligned}\ea

Using the optimal condition $\delta\theta=0$, we have $2\delta\sigma_{ab}=\mathcal{L}_{\xi}\gamma_{ab}$. Then, we can get
\ba\begin{aligned}
  \math{E}_\S(g,\d g)=-4\int_{\Sigma}\tilde{\bm\epsilon}\delta \s_{bc}A^{bc}+\math{O}(\e)\,.
\end{aligned}\ea
with
\ba\begin{aligned}
&A_{ij}=P^{z}{}_{ij}{}^{def}\nabla_d\delta g_{ef}\\
&=\d g_{ui;j}-\frac{1}{2}\d g_{ij;u}=\d g_{ui;j}-\d \s_{ij}
\end{aligned}\ea
Using the Gaussian null coordinate, it is not hard to verify $\d g_{ui;j}=0$. Therefore, we have $A_{ij}=-\d \s_{ij}$, i.e.
\ba\begin{aligned}
\math{E}_\S(g,\d g)=4\int_{\Sigma}\tilde{\bm\epsilon}\delta\sigma_{bc}\delta\sigma^{bc}+\math{O}(\e)\geq \math{O}(\e)\,.
\end{aligned}\ea

Therefore, the second-order perturbation inequality is
\ba\begin{aligned}
\delta^2 f(r_h)\leq \mathcal{O}(\epsilon,\lambda)\,.
\end{aligned}\ea
After considering the first-order optimal condition, \eqref{flambda} becomes
\ba\begin{aligned}\label{secondflambda}
f(\lambda)=&-\epsilon\lambda r_h\delta f'(r_h)-\frac{1}{2}\epsilon^2 r_h^2 f''(r_h)\\
&+\frac{1}{2}\lambda^2\left[\delta^2f(r_h)+2\delta r_m\delta f'(r_h)+\delta r_m^2f''(r_h)\right]\,.
\end{aligned}\ea
Considering $r_m(\lambda)$ is the minimum value of $f(r_m(\lambda),\lambda)$, we have
\ba\begin{aligned}
\delta r_m=\frac{\delta f'(r_m)}{f''(r_m)}=\frac{\delta f'(r_h)}{f''(r_h)}+\math{O}(\e)\,,
\end{aligned}\ea
and therefore, with the identity $\delta^2f(r_h)\leq \mathcal{O}(\e)$, we can find \eqref{secondflambda} can be further expressed as
\ba\begin{aligned}
  f(\lambda)=-\frac{\left[\lambda\delta f'(r_h)+\epsilon r_h f''(r_h)\right]^2}{2f''(r_h)}\,,
\end{aligned}\ea
under the second-order approximation of $\epsilon$ and $\lambda$. Therefore, we have $f(\lambda)\leq 0$ under the second-order approximation, which means the black hole cannot be destroyed.

\section{Conclusion and Discussion}\label{sec6}

Apart from the KN black holes, the hairy black holes in general relativity have been widely investigated in gravity and cosmology. In the current paper, we extended the Sorce-Wald gedanken experiment to study the WCCC of the static and spherically symmetric hairy black holes in Einstein gravity coupled to some matter fields. We considered a collision process without requiring the spherical symmetry of the perturbation matter fields. After assuming the stability condition of the spacetime and based on the Noether charge method by Iyer and Wald \cite{Iyer:1994ys}, we derived the first- and second-order perturbation inequalities which reflect the null energy condition of the matter fields. Using these inequalities, we found that the nearly extremal hairy black holes cannot be destroyed under the second-order approximation of perturbation and therefore the WCCC is valid in these hairy black holes. Our result is universal and independent of the explicit expressions of spacetime metric and the Lagrangian for matter fields. Therefore, it implies that the WCCC is generally valid for any black holes in Einstein gravity as long as the matter fields satisfy the null energy condition.

In this paper, we only considered the cases where the background spacetime is static and spherically symmetric. To give general proof of the WCCC in general relativity under the perturbation level, it is necessary to consider the perturbed spinning black holes in which the perturbation matter fields carry the total angular momentum. Moreover, it is also interesting to extend the discussion into the modified gravitational theories, such as the Einstein-scalar-Gauss-Bonnet gravity and Lovelock gravities.

\section*{Acknowledgement}
This article is supported by the GuangDong Basic and Applied Basic Research Foundation with Grant no. 217200003 and the Talents Introduction Foundation of Beijing Normal University with Grant no. 310432102.

\appendix

\section{Proof 1}
In this appendix, we would like to proof that the first-order optimal condition at an arbitrary hypersurface $H_1$, which is the horizon on the background, is equivalent to $\delta T_{kk}=0$ at a null hypersurface $H_2$.

We assume the first order optimal condition is satisfied,i.e.
\ba\begin{aligned}
%\left.\delta \left[T_{a}^b (d\tilde{z})_b\left(\frac{\partial}{\partial \tilde{u}}\right)^a\right] \right|_{H_1}=0\,,
\delta T_u^z=0
\end{aligned}\ea
where $T_u^z=T_u^z(\tilde{x},\lambda)$ and we use $O_1:\{\tilde{u},\tilde{z},\tilde{\theta}\}$ to represent the coordinate system that we used to calculate the first-order inequality and $H_1$ is the hypersurface given by $\tilde{z}=0$. We will use ``$\sim$'' to label the quantities valued in $O_1$ later. Then, we assume $O_2:\{u,z,\theta\}$ is the Gaussian null coordinate around the null hypersurface $H_2$ with $k^a=\left({\partial}/{\partial u}\right)^a$ the tangent null vector of $H_2$. Taking a point $p_1$ on $H_1$ with coordinate $\{\tilde{x}_{p1}\}$ and a point $p_2$ on $H_2$ with coordinate $\{\tilde{x}_{p2}\}$ in $O_1$, here we use ``$\sim$'' to represent the coordinate value in $O_1$. We also assume that the coordinate of $p_2$ in $O_2$ is $\{\hat{x}_{p2}\}$. We associate $p_1$ and $p_2$ by demanding $\hat{x}_{p2}=\tilde{x}_{p1}$. In $O_1$, the stress energy tensor can be expressed as
\begin{equation}\label{energytensorapp}
  T_{ab}(\lambda)=\tilde{T}_{\mu\nu}(\tilde{x},\lambda)(d\tilde{x}^{\mu})_a(d\tilde{x}^{\nu})_b\,.
\end{equation}
The optimal condition valued at $p_1$ gives
\be\begin{aligned}\label{optimalapp}
\delta \tilde{T}_0^1(\tilde{x}_{p1})=\delta\tilde{T}_0^1(\hat{x}_{p2})=0\,.
\end{aligned}\ee
%Noting that the variation in \eqref{optimalapp} is evaluated in $O_1$ and therefore the coordinate basis $\left(\frac{\partial}{\partial \tilde{x}^\mu}\right)^a$ and the dual basis $(d\tilde{x}^\mu)_a$ is fixed.
% This condition is hold for any $p_2$ on $H_2$, therefore we denote $x_{p2}$ as $x$ for simplicity.
Then, in the null hypersurface,
\be\begin{aligned}\label{nullenergy}
  T_{ab}k^ak^b|_{p_2}&=\left.T_{ab}\left(\frac{\partial}{\partial u}\right)^a \left(\frac{\partial}{\partial u}\right)^b\right|_{p2}\\
  &=\tilde{T}_{\mu\nu}(\tilde{x}_{p2},\lambda)\left.\frac{\partial\tilde{x}^{\mu}}{\partial u}\right|_{p2}\left.\frac{\partial\tilde{x}^{\nu}}{\partial u}\right|_{p2}\,.
\end{aligned}\ee

Then, we would like to calculate the variation of Eq. \eqref{nullenergy} in $O_2$. We assume the coordinates transformation between $O_1$ and $O_2$ is $\tilde{x}_q=\tilde{x}(\hat{x}_q,\lambda)$ for any point $q$ and the coordinates transformation satisfies $\tilde{x}(\hat{x},0)=\hat{x}=x$. Since the equation \eqref{nullenergy} holds for any $p_2$ at $H_2$, therefore we will get rid of $p_2$ for simplicity in the following calculation. Therefore \eqref{nullenergy} can be written as
\ba\begin{aligned}
T_{kk}(\hat{x},\lambda)=\tilde{T}_{\mu\nu}(\tilde{x}(\hat{x},\lambda),\lambda)\frac{\partial\tilde{x}^{\mu}(\hat{x},\lambda)}{\partial u}\frac{\partial\tilde{x}^{\nu}(\hat{x},\lambda)}{\partial u}\,.
\end{aligned}\ea
When we calculate the variation, we fix $\hat{x}$ both side. Then, the variation of $T_{kk}(\hat{x},\lambda)$ will be
\ba\begin{aligned}\label{partiallambda}
 \delta T_{kk}=&[\delta\tilde{T}_{\mu\nu}(x)+\partial_{\rho}\tilde{T}_{\mu\nu}(x)\delta\tilde{x}^{\rho}(x)]\frac{\partial \tilde{x}^{\mu}(\hat{x},0)}{\partial u}\frac{\partial \tilde{x}^{\nu}(\hat{x},0)}{\partial u}\\
& +2\tilde{T}_{\mu\nu}(x)\frac{\partial \tilde{x}^{\mu}(\hat{x},0)}{\partial u}\frac{\partial\delta\tilde{x}^{\nu}(x)}{\partial u}\,.
\end{aligned}\ea
%&=\partial_{\lambda}\tilde{T}_{00}(\tilde{x}(x,\lambda),\lambda)\mid_{\lambda=0}+2\tilde{T}_{0\nu}(x)\frac{\partial \delta \tilde{x}^{\nu}}{\partial v}\\
Considering the coordinates of a point is the same in $O_1$ and $O_2$ when $\lambda=0$, we can find
\ba
\frac{\partial \tilde{x}^{\mu}(\hat{x},\lambda)}{\partial u}|_{\lambda=0}=\delta_0^\mu\,.
\ea
Then, considering $T_{0i}$ components vanish on background and $T_{00}=0$ on the horizon $z=0$, we can obtain
\ba\begin{aligned}\label{deltatkk}
 \delta T_{kk}
 %&=\left.\frac{d\tilde{T}_{00}(\tilde{x}(\hat{x},\lambda),\lambda)}{d\lambda}\right|_{\lambda=0}+2\tilde{T}_{01}(\hat{x})\left.\frac{\partial \pd_{\lambda} \tilde{x}^1(\hat{x},\lambda)}{\partial u}\right|_{\lambda=0}\\
 =\delta\tilde{T}_{00}(x)+\partial_{1} \tilde{T}_{00}(x) \delta\tilde{x}^{1}(x) +2\tilde{T}_{01}({x})\frac{\partial \delta \tilde{x}^1({x})}{\partial u}\,.\quad
\end{aligned}\ea
Using the fact that $g_{ab}(\frac{\partial}{\partial u})^a(\frac{\partial}{\partial u})^b|_{H2}=0$, with a similar calculation, we can find
\ba\begin{aligned}
&\delta g_{uu}=\delta\tilde{g}_{00}(x)+\partial_{1} \tilde{g}_{00}(x) \delta\tilde{x}^{1}(x) +2\tilde{g}_{01}({x})\frac{\partial \delta \tilde{x}^1({x})}{\partial u}=0\,,\\
  %&\partial_{\lambda}\tilde{g}_{00}(\hat{x},\lambda)|_{\lambda=0}+\partial_{1} \tilde{g}_{00}\partial_{\lambda}\tilde{x}^{1}(\hat{x},\lambda)|_{\lambda=0} +\left.\frac{\partial \pd_{\lambda} \tilde{x}^1(\hat{x},\lambda)}{\partial u}\right|_{\lambda=0}=0\\
 %&\delta\tilde{g}_{00}(x,\lambda)+\partial_{1} \tilde{g}_{00}\partial_{\lambda}\tilde{x}^{1} +2\tilde{g}_{01}(x)\frac{\partial \delta \tilde{x}^{1}}{\partial u}=0\\
&2\frac{\partial \delta \tilde{x}^{1}}{\partial u}=-\delta\tilde{g}_{00}(x)-\partial_{1} \tilde{g}_{00}\delta\tilde{x}^{1}\,.
\end{aligned}\ea
The zero component of the conservation law $\nabla_a T^{ab}=0$ gives
\ba\begin{aligned}
g^{ac}\partial_c T_{a0}-\Gamma^{da}{}_a T_{d0}-\Gamma^{da}{}_0 T_{ad}=0\,.
\end{aligned}\ea
On the background, the line element can be expressed as
\ba\begin{aligned}
ds^2=2 (dz-z\alpha du-z\beta_i d\theta^i)du+\g_{ij} d\q^id\q^j\,.
\end{aligned}\ea
It is early to find
\ba\begin{aligned}
g^{ac}\Gamma^1{}_{ac}=-2\alpha\,,\quad \Gamma^{\mu\nu}{}_0=- \Gamma^{\nu\mu}{}_0\,.
\end{aligned}\ea
Therefore, we have
\ba\begin{aligned}
\partial_1 T_{00}+2\alpha T_{01}=0\,,
\end{aligned}\ea
i.e.
\ba\begin{aligned}
\partial_{1} {T}_{00}-{T}_{01}\partial_{1} {g}_{00}=0\,.
\end{aligned}\ea
Then, we have
\ba\begin{aligned}
 \delta T_{kk}&=\delta\tilde{T}_{00}-\tilde{T}_{01}\partial_{1}\tilde{g}_{00}\delta\tilde{x}^1+\tilde{T}_{01}(-\delta\tilde{g}_{00}-\partial_1\tilde{g}_{00}\delta\tilde{x}^1)\\
 &=\delta\tilde{T}_{00}-\tilde{T}_{01}\delta\tilde{g}_{00}\\
 &=\delta\tilde{T}^1_0=0,
\end{aligned}\ea
which is the optimal condition on the null hypersurface $H_2$.

\section{Proof 2}

In this appendix, we would like to proof that there exists a null hypersurface linking $\tilde{r}=r_h$ in $O_1:\{\tilde{v},\tilde{r},\tilde{\theta}^i\}$ and $r=r_h$ in $O_2:\{v,r,\theta^i\}$ for an extremal black hole when the optimal condition is satisfied.

We assume a null hypersurface $\Sigma$ expressed as $\tilde{r}=r_h$ at early time and is the horizon on background. Assuming there is a tangent vector $\left(\frac{\partial}{\partial u}\right)^a$ of $\Sigma$ and the Gaussian null coordinate of $\Sigma$ is $O_\text{GN}:\{u,z,\tilde{\theta}^i\}$, then, the expression of $\Sigma$ in $O_2$ is
\ba\begin{aligned}
v=u\,,\quad r=r(u,\tilde{\theta}^i,\lambda)\,,\quad \theta^i=\theta^i(u,\tilde{\theta}^i,\lambda)\,,
\end{aligned}\ea
where $r(u,\tilde{\theta}^i,0)=r_h\,,\theta^i=\theta^i(u,\tilde{\theta}^i,0)=\tilde{\theta}^i$. We can get the coordinate basis in Gaussian null coordinate system
\ba\begin{aligned}
&k^a=\left(\frac{\partial}{\partial u}\right)^a=\left(\frac{\partial}{\partial v}\right)^a+\frac{\pd r}{\pd u}\left(\frac{\partial}{\partial r}\right)^a+\frac{\pd \theta^i}{\pd u}\left(\frac{\partial}{\partial\theta^i}\right)^a\,,\\
&(e_i)^a=\left(\frac{\partial}{\partial \tilde{\theta}^i}\right)^a=\frac{\pd r}{\pd \tilde{\theta}^i}\left(\frac{\partial}{\partial r}\right)^a+\frac{\pd\theta^j}{\pd\tilde{\theta}^i}\left(\frac{\partial}{\partial\theta^j}\right)^a\,.
\end{aligned}\ea
Since the line element in $O_2$ is
\ba\begin{aligned}
ds^2=-f(r,\lambda)dv^2+2\chi(r,\lambda)dvdr+r^2h_{ij} d\theta^i d\theta^j\,,
\end{aligned}\ea
and considering the Gaussian null coordinate of a null hypersurface satisfies
\ba\begin{aligned}
k^ak_a=0\,,\quad k^a(e_i)_a=0\,,
\end{aligned}\ea
we can find
\ba\begin{aligned}\label{kappendixb}
%&k^a k_a=-f(r)+2\chi(r)\frac{\partial r}{\partial u}+r^2 h_{ij}\frac{\pd \theta^i}{\pd u}\frac{\pd \theta^j}{\pd u}=0\\
%&k^a(e_i)_a=2\chi(r)\frac{\pd r}{\pd \theta^i}+r^2h_{jl}\frac{\pd \theta^l}{\pd u}\frac{\pd \theta^j}{\pd \tilde{\theta}^i}=0\\
&k^a k_a=-f(r,\lambda)+2\chi(r,\lambda)k^r+r^2 h_{ij}k^ik^j=0\,,\\
&k^a(e_i)_a=2\chi(r,\lambda)(e_i)^r+r^2h_{jl}k^l(e_i)^j=0\,.
\end{aligned}\ea
Considering $k^r=k^i=(e_i)^r=0$ and $(e_i)^j=\delta_i^j$ on the background geometry $\l=0$, through the variation of \eqref{kappendixb}, we can obtain
\ba\begin{aligned}
\delta k^r&=\frac{1}{2\chi(r_h)}\left.\frac{d}{d\lambda}f\left(r(u,\tilde{\theta}^i,\lambda\right),\lambda)\right|_{\lambda=0}\\
&=\frac{1}{2\chi(r_h)}[\partial_r f(r_h) \delta r+\delta f]\,,\\
\delta k^j&=-\frac{2}{r_h}h^{ij}\chi(r_h)\delta(e_i)^r\,.
\end{aligned}\ea
Noting that the variation is done fixing $\{u,z,\tilde{\theta}^i\}$. For an extremal black hole, we have $\partial_r f(r_h)=0$. When the first-order optimal condition is satisfied,i.e. $\delta f=0$, we can find $\delta k^r=0$, which gives
\ba
\frac{\partial\delta r}{\partial u}=0\,.\\
\ea
This means that $\delta r$ is a constant along $k^a$.

Then, according to the discussion in appendix B, it is obvious that the optimal condition demands that $\d T_{ab}k^a k^b=0$. From the Raychaudhuri equation on $\Sigma$, we have
\ba\begin{aligned}\label{raychaudhuri}
\frac{d\vartheta(\l)}{du}=&-\frac{1}{n-2}\vartheta(\l)^2\\
&-\sigma_{ab}(\l)\sigma^{ab}(\l)-T_{ab}(\l)k^a k^b\,,
\end{aligned}\ea
where
\ba\begin{aligned}
\vartheta&=\frac{1}{2}\g^{ab}\mathcal{L}_k \g_{ab}\\
\sigma_{ab}&=\frac{1}{2}\mathcal{L}_k \g_{ab}-\frac{1}{n-2}\g_{ab}\vartheta\,,
\end{aligned}\ea
Then, considering the fact that $\vartheta(0)=0\,,\sigma_{ab}(0)=0$ on background, the optimal condition $\delta T_{kk}=0$ implies
\ba\begin{aligned}
\frac{d\delta\vartheta}{du}=0\,,
\end{aligned}\ea
which implies $\delta\vartheta=0$ on the entire horizon since $\delta\vartheta=0$ at early-time cross-section $B_0$.

Since $\gamma_{ab}$ is independent with $u$ on the background, the variation of $\vartheta$ can be written as
\ba\begin{aligned}\label{expansionappb}
\delta\vartheta=\frac{1}{2}\gamma^{ij}\mathcal{L}_k\delta\gamma_{ij}\,,
\end{aligned}\ea
where $\gamma_{ij}$ is the metric component in $O_\text{GN}$. Next, we would like to calculate $\delta \vartheta$. Using the coordinate transformation relation, we can express $\gamma_{ij}$ using the line element in $O_2$
\ba\begin{aligned}
  \gamma_{ij}(\tilde{x},\lambda)=\frac{\pd x^{\mu}(\tilde{x},\lambda)}{\pd \tilde{\theta}^i}\frac{\pd x^{\nu}(\tilde{x},\lambda)}{\pd\tilde{\theta}^j}g_{\mu\nu}(x(\tilde{x},\lambda),\lambda)\,,
\end{aligned}\ea
where we used $\tilde{x}$ to denote the coordinates in $O_\text{GN}$. When $\lambda=0$, there will be
\ba\begin{aligned}
\gamma_{ij}(\tilde{x})=r^2h_{ij}(x(\tilde{x},0),0)=r^2h_{ij}(\tilde{x})=\tilde{h}_{ij}(\tilde{x})
\end{aligned}\ea
and the variation of $\gamma_{ij}$ will be
\ba\begin{aligned}
\delta\gamma_{ij}(\tilde{x})&=\left.\frac{d}{d\lambda}\left[\frac{\pd x^{\mu}(\tilde{x},\lambda)}{\pd \tilde{\theta}^i}\frac{\pd x^{\nu}(\tilde{x},\lambda)}{\pd\tilde{\theta}^j}g_{\mu\nu}(x(\tilde{x},\lambda),\lambda)\right]\right|_{\lambda=0}\\
&=2\frac{\pd\delta x^{\mu}(\tilde{x})}{\pd \tilde{\theta}^i}\frac{\pd x^{\nu}(\tilde{x})}{\pd\tilde{\theta}^j}g_{\mu\nu}(\tilde{x})\\
&\quad+\frac{\pd x^{\mu}(\tilde{x})}{\pd \tilde{\theta}^i}\frac{\pd x^{\nu}(\tilde{x})}{\pd\tilde{\theta}^j}(\delta g_{\mu\nu}(\tilde{x})+\partial_{\sigma}g_{\mu\nu}(\tilde{x})\delta x^{\sigma})\\
&=2\frac{\pd\delta x^{l}(\tilde{x})}{\pd \tilde{\theta}^i}\tilde{h}_{l j}(\tilde{x})+\delta x^{\sigma}\partial_{\sigma}\tilde{h}_{ij}(\tilde{x})\,,
\end{aligned}\ea
%After evaluating it at $\lambda=0$, we can find
%\ba\begin{aligned}
%  \delta\gamma_{ij}(\tilde{x})&=2\frac{\pd\delta x^{\mu}(\tilde{x})}{\pd \tilde{\theta}^i}g_{\mu j}(\tilde{x})+\frac{d}{d\lambda}g_{ij}(x(\tilde{x},\lambda),\lambda)\\
%  &=2\frac{\pd\delta x^{l}(\tilde{x})}{\pd \tilde{\theta}^i}\tilde{h}_{l j}(\tilde{x})+\delta x^{\sigma}(\tilde{x})\partial_{\sigma}\tilde{h}_{ij}(\tilde{x})+\delta \tilde{h}_{ij}(\tilde{x})\\
%  &=2\frac{\pd\delta x^{l}(\tilde{x})}{\pd \tilde{\theta}^i}\tilde{h}_{l j}(\tilde{x})+\delta x^{\sigma}\partial_{\sigma}\tilde{h}_{ij}(\tilde{x})\,,
%\end{aligned}\ea
where we used $\tilde{h}_{ij}$ is independent with $\lambda$. And we can further obtain
\ba\begin{aligned}
\mathcal{L}_k\delta\gamma_{ij}&=\frac{\partial}{\partial u}\delta\gamma_{ij}=2\frac{\partial\delta\partial_0 x^l}{\partial\tilde{\theta}^i}\tilde{h}_{lj}+\delta(\partial_0 x^{\sigma})\partial_{\sigma}\tilde{h}_{ij}\\
&=2\frac{\partial\delta k^l}{\partial\tilde{\theta}^i}\tilde{h}_{lj}+\delta k^r\partial_r \tilde{h}_{ij}+\delta k^l \partial_l \tilde{h}_{ij}\\
&=2\frac{\partial\delta k^l}{\partial\tilde{\theta}^i}\tilde{h}_{lj}+\delta k^l \partial_l \tilde{h}_{ij}
\end{aligned}\ea

Then, \eqref{expansionappb} becomes
\ba\begin{aligned}\label{optimalappb}
\delta\vartheta&=\frac{1}{2}\gamma^{ij}\mathcal{L}_k\delta\gamma_{ij}\\
&=\frac{\partial\delta k^l}{\partial\tilde{\theta}^i}\tilde{h}_{lj}\tilde{h}^{ij}+\frac{1}{2}\delta k^l\tilde{h}^{ij} \partial_l \tilde{h}_{ij}\\
&=2\frac{1}{\sqrt{\gamma}}\partial_i(\sqrt{\gamma}\delta k^i)
\end{aligned}\ea
Thus, the optimal condition gives
\ba\begin{aligned}
\delta\vartheta&=D_i\delta k^i=D_i[-\frac{2}{r_h^2}h^{ij}\chi(r_h)\delta(e_j)^r]\\
&=-\frac{2}{r_h^2}\chi D_a D^a \delta r=0\,,
\end{aligned}\ea
where we use $D_i$ to denote the derivative operator adopted to $\tilde{h}_{ij}$. Therefore, we find that the optimal condition gives $D_a D^a\delta r=0$. This is condition also gives
\ba\begin{aligned}
0&=\int_{B}d^{(n-2)}\theta\sqrt{\gamma}\delta r D_aD^a\delta r\\
&=\int_{B}d^{(n-2)}\theta\sqrt{\gamma} D_a(\delta r D^a\delta r)-\int_{B}d^{(n-2)}\theta\sqrt{\gamma} D_a\delta r D^a\delta r\,.
\end{aligned}\ea
Since the sphere with no boundary, the boundary term vanishes. Considering $D^a \delta r$ is spacelike, we can find
\ba\begin{aligned}
  D^a \delta r=0\,,
\end{aligned}\ea
which means $\delta r$ is a constant. Thus, the hypersurface at late time under the first-order approximation is given by $r=constant$. Considering $\delta\vartheta=0$, the area of the cross-section is a constant under the variation. Therefore, we find the null hypersurface which is $r=r_h$ at late time.
\\
\\


\begin{thebibliography}{100}
\bibitem{Penrose:1969pc}
R.~Penrose, Gravitational collapse: The role of general relativity, Riv. Nuovo Cim. \textbf{1} , 252-276 (1969).

%\cite{Wald:1997wa}
\bibitem{Wald:1997wa}
R.~M.~Wald,
%``Gravitational collapse and cosmic censorship,''
doi:10.1007/978-94-017-0934-7\_5
[arXiv:gr-qc/9710068 [gr-qc]].
%282 citations counted in INSPIRE as of 04 Jul 2021

\bibitem{Wald:1974wl}
R. Wald, Gedanken experiments to destroy a black hole,
Ann. Phys. (N.Y.) {\bf82}, 548 (1974).

%\cite{Hubeny:1998ga}
\bibitem{Hubeny:1998ga}
V.~E.~Hubeny,
%``Overcharging a black hole and cosmic censorship,''
Phys. Rev. D \textbf{59}, 064013 (1999)
doi:10.1103/PhysRevD.59.064013
[arXiv:gr-qc/9808043 [gr-qc]].
%186 citations counted in INSPIRE as of 04 Jul 2021

%\cite{Jacobson:2010iu}
\bibitem{Jacobson:2010iu}
T.~Jacobson and T.~P.~Sotiriou,
%``Destroying black holes with test bodies,''
J. Phys. Conf. Ser. \textbf{222}, 012041 (2010)
doi:10.1088/1742-6596/222/1/012041
[arXiv:1006.1764 [gr-qc]].
%26 citations counted in INSPIRE as of 04 Jul 2021

%\cite{Chirco:2010rq}
\bibitem{Chirco:2010rq}
G.~Chirco, S.~Liberati and T.~P.~Sotiriou,
%``Gedanken experiments on nearly extremal black holes and the Third Law,''
Phys. Rev. D \textbf{82}, 104015 (2010)
doi:10.1103/PhysRevD.82.104015
[arXiv:1006.3655 [gr-qc]].
%36 citations counted in INSPIRE as of 04 Jul 2021

%\cite{Saa:2011wq}
\bibitem{Saa:2011wq}
A.~Saa and R.~Santarelli,
%``Destroying a near-extremal Kerr-Newman black hole,''
Phys. Rev. D \textbf{84}, 027501 (2011)
doi:10.1103/PhysRevD.84.027501
[arXiv:1105.3950 [gr-qc]].
%88 citations counted in INSPIRE as of 04 Jul 2021

%\cite{Gao:2012ca}
\bibitem{Gao:2012ca}
S.~Gao and Y.~Zhang, ``Destroying extremal Kerr-Newman black holes with test particles,''
Phys. Rev. D \textbf{87}, no.4, 044028 (2013)
doi:10.1103/PhysRevD.87.044028
[arXiv:1211.2631 [gr-qc]].
%93 citations counted in INSPIRE as of 04 Jul 2021

%N
%\cite{Zimmerman:2012zu}
\bibitem{Zimmerman:2012zu}
P.~Zimmerman, I.~Vega, E.~Poisson and R.~Haas,
%``Self-force as a cosmic censor,''
Phys. Rev. D \textbf{87}, no.4, 041501 (2013)
doi:10.1103/PhysRevD.87.041501
[arXiv:1211.3889 [gr-qc]].
%71 citations counted in INSPIRE as of 04 Jul 2021
%\cite{Barausse:2010ka}
\bibitem{Barausse:2010ka}
E.~Barausse, V.~Cardoso and G.~Khanna,
%``Test bodies and naked singularities: Is the self-force the cosmic censor?,''
Phys. Rev. Lett. \textbf{105}, 261102 (2010)
doi:10.1103/PhysRevLett.105.261102
[arXiv:1008.5159 [gr-qc]].
%153 citations counted in INSPIRE as of 04 Jul 2021
Barausse:2010ka
%\cite{Barausse:2011vx}
\bibitem{Barausse:2011vx}
E.~Barausse, V.~Cardoso and G.~Khanna,
%``Testing the Cosmic Censorship Conjecture with point particles: the effect of radiation reaction and the self-force,''
Phys. Rev. D \textbf{84}, 104006 (2011)
doi:10.1103/PhysRevD.84.104006
[arXiv:1106.1692 [gr-qc]].
%91 citations counted in INSPIRE as of 04 Jul 2021

%\cite{Colleoni:2015ena}
\bibitem{Colleoni:2015ena}
M.~Colleoni, L.~Barack, A.~G.~Shah and M.~van de Meent,
%``Self-force as a cosmic censor in the Kerr overspinning problem,''
Phys. Rev. D \textbf{92}, no.8, 084044 (2015)
doi:10.1103/PhysRevD.92.084044
[arXiv:1508.04031 [gr-qc]].
%74 citations counted in INSPIRE as of 04 Jul 2021

%\cite{Sorce:2017dst}
\bibitem{Sorce:2017dst}
J.~Sorce and R.~M.~Wald,
%``Gedanken experiments to destroy a black hole. II. Kerr-Newman black holes cannot be overcharged or overspun,''
Phys. Rev. D \textbf{96}, no.10, 104014 (2017)
doi:10.1103/PhysRevD.96.104014
[arXiv:1707.05862 [gr-qc]].
%107 citations counted in INSPIRE as of 04 Jul 2021


\bibitem{Israel:1967wq}
W.~Israel, ``Event horizons in static vacuum space-times,'' Phys. Rev. \textbf{164}, 1776-1779 (1967).
\bibitem{Carter:1971zc}
B.~Carter, ``Axisymmetric Black Hole Has Only Two Degrees of Freedom,'' Phys. Rev. Lett. \textbf{26}, 331-333 (1971).
\bibitem{Ruffini:1971bza}
R.~Ruffini and J.~A.~Wheeler, ``Introducing the black hole,''
Phys. Today \textbf{24}, no.1, 30 (1971)
\bibitem{MSDV}
M. S. Volkov and D. V. Galtsov, ``Non-Abelian Einstein Yang-Mills black holes'', JETP Lett. \textbf{50}, 346 (1989).
\bibitem{Bizon:1990sr}
P.~Bizon, ``Colored black holes,'' Phys. Rev. Lett. \textbf{64}, 2844-2847 (1990).
\bibitem{Greene:1992fw}
B.~R.~Greene, S.~D.~Mathur and C.~M.~O'Neill, ``Eluding the no hair conjecture: Black holes in spontaneously broken gauge theories,''
Phys. Rev. D \textbf{47}, 2242-2259 (1993).
\bibitem{Kanti:1995vq}
P.~Kanti, N.~E.~Mavromatos, J.~Rizos, K.~Tamvakis and E.~Winstanley, ``Dilatonic black holes in higher curvature string gravity,''
Phys. Rev. D \textbf{54}, 5049-5058 (1996).
\bibitem{Luckock:1986tr}
H.~Luckock and I.~Moss, ``BLACK HOLES HAVE SKYRMION HAIR,'' Phys. Lett. B \textbf{176}, 341-345 (1986).
\bibitem{Droz:1991cx}
S.~Droz, M.~Heusler and N.~Straumann, ``New black hole solutions with hair,''
Phys. Lett. B \textbf{268}, 371-376 (1991).
% some reference
\bibitem{Herdeiro1}
 C. Herdeiro and E. Radu, Phys. Rev. Lett. {\bf112}, 221101
(2014).
\bibitem{Herdeiro2}
C. Herdeiro and E. Radu, Classical Quantum Gravity {\bf32},
144001 (2015).
\bibitem{Herdeiro:2015waa}
C.~A.~R.~Herdeiro and E.~Radu, ``Asymptotically flat black holes with scalar hair: a review,''
Int. J. Mod. Phys. D \textbf{24}, no.09, 1542014 (2015).
%\cite{Planck:2015fie}
\bibitem{Hong:2020miv}
J.~P.~Hong, M.~Suzuki and M.~Yamada, ``Spherically Symmetric Scalar Hair for Charged Black Holes,''
Phys. Rev. Lett. \textbf{125}, no.11, 111104 (2020).


\bibitem{Planck:2015fie}
P.~A.~R.~Ade \textit{et al.} [Planck],
%``Planck 2015 results. XIII. Cosmological parameters,''
Astron. Astrophys. \textbf{594}, A13 (2016)
doi:10.1051/0004-6361/201525830
[arXiv:1502.01589 [astro-ph.CO]].
%9932 citations counted in INSPIRE as of 04 Jul 2021

%\cite{Bertone:2004pz}
\bibitem{Bertone:2004pz}
G.~Bertone, D.~Hooper and J.~Silk,
%``Particle dark matter: Evidence, candidates and constraints,''
Phys. Rept. \textbf{405}, 279-390 (2005)
doi:10.1016/j.physrep.2004.08.031
[arXiv:hep-ph/0404175 [hep-ph]].
%4195 citations counted in INSPIRE as of 04 Jul 2021

%\cite{Peebles:2002gy}
\bibitem{Peebles:2002gy}
P.~J.~E.~Peebles and B.~Ratra,
%``The Cosmological Constant and Dark Energy,''
Rev. Mod. Phys. \textbf{75}, 559-606 (2003)
doi:10.1103/RevModPhys.75.559
[arXiv:astro-ph/0207347 [astro-ph]].
%3872 citations counted in INSPIRE as of 04 Jul 2021

%\cite{Hollands:2012sf}
\bibitem{Hollands:2012sf}
S.~Hollands and R.~M.~Wald,
%``Stability of Black Holes and Black Branes,''
Commun. Math. Phys. \textbf{321}, 629-680 (2013)
doi:10.1007/s00220-012-1638-1
[arXiv:1201.0463 [gr-qc]].
%124 citations counted in INSPIRE as of 04 Jul 2021

%\cite{Iyer:1994ys}
\bibitem{Iyer:1994ys}
V.~Iyer and R.~M.~Wald,
%``Some properties of Noether charge and a proposal for dynamical black hole entropy,''
Phys. Rev. D \textbf{50}, 846-864 (1994)
doi:10.1103/PhysRevD.50.846
[arXiv:gr-qc/9403028 [gr-qc]].
%1526 citations counted in INSPIRE as of 04 Jul 2021

\end{thebibliography}
\end{document}